\documentclass[journal=jpclcd, manuscript=letter]{achemso}
\setkeys{acs}{articletitle = true, chaptertitle = true, doi = true}

\usepackage[version=3]{mhchem} 
\usepackage[T1]{fontenc}       
\usepackage{mathptmx}

\usepackage{graphicx}  
\usepackage{bm}        
\usepackage{bbm}       
\usepackage{amssymb}   
\usepackage{amsmath}   
\usepackage{physics}   
\usepackage{textgreek} 
\usepackage{multirow}  
\usepackage{booktabs}  
\usepackage{hyperref}  
\hypersetup{colorlinks=true, linkcolor=black, urlcolor=black, citecolor=black, plainpages=false,
bookmarksnumbered=true}
\usepackage{xurl}      
\usepackage{siunitx}   
\usepackage{xcolor}    
\usepackage[utf8]{inputenc} 
\author{Christof Holzer}
\affiliation{Institute of Theoretical Solid State Physics, Karlsruhe
Institute of Technology (KIT), Wolfgang-Gaede-Stra\ss{}e 1, 76131 Karlsruhe,
Germany}
\email{christof.holzer@kit.edu}

\author{Yannick J. Franzke}
\affiliation{Institute of Nanotechnology, Karlsruhe
Institute of Technology (KIT), Kaiserstra\ss{}e 12, 76131 Karlsruhe, Germany}
\email{yannick.franzke@kit.edu}

\title{A Guide to Molecular Properties from the Bethe--Salpeter Equation}

\begin{document}

\renewcommand*\tocentryname{ }
\setlength{\fboxrule}{0 pt}
\begin{tocentry}
\begin{center}
\includegraphics[scale=1.0]{TOC_blank.eps}
\end{center}
\end{tocentry}


\begin{abstract}
The Bethe--Salpeter equation (BSE) combined with the Green's function
$GW$ method has successfully transformed into a robust computational
tool to describe light-matter interactions and excitation spectra for
molecules, solids, and materials from first principles.
Thanks to its ability to accurately describe charge-transfer and Rydberg
excitations, the $GW$-BSE already forms an established and cost-efficient
alternative to time-dependent density functional theory. This raises the
question whether the $GW$-BSE approach can become a more general framework
for molecular properties beyond excitation energies.
In this mini-review, we recapitulate recent endeavors along this point
in terms of both theoretical and practical developments for quantum
chemistry, physical chemistry, and related fields.
In doing so, we provide guidelines for current applications to
chemical challenges in collaboration with experimentalists as well as
to future developments to extended the $GW$-BSE toolkit.
\end{abstract}

%
%
\section{Introduction}
\label{sec:introduction}

The Bethe--Salpeter equation (BSE) method in the Green's function $GW$ approximation
has seen significant success for predicting optical and excitonic spectra in
theoretical solid state physics and materials science.
In the recent years, the $GW$-BSE also successfully made its transition 
towards theoretical chemistry,\cite{Blase.Duchemin.ea:Bethe-Salpeter.2020}
becoming a powerful tool to model light-matter interactions
of various molecular systems.
Traditionally, these interactions are studied with time-dependent density
functional theory (TD-DFT). Here, a linear response equation consisting of
the Kohn--Sham orbital eigenvalues and the electronic Hessian, including
the two-electron Coulomb and exact exchange interaction as well as the exchange-correlation
(XC) kernel, is solved to describe the excited states. \cite{Casida:Time-dependent.2009}
This framework was applied with great success over a wide range of molecular systems,
however, the accuracy strongly depends on the chosen functional approximation.
Compared to ground-state DFT calculations, it is generally more difficult to find
the ``right'' functional for TD-DFT, as different types of excitations such as
charge-transfer (CT), Rydberg, spin-flip, or two-photon excitations may require
different functional approximations.

The theoretical foundation of the BSE approach is completely different than
that of DFT or TD-DFT, as it is not based on the electron density but
on the one-particle Green’s function in a many-body perturbation theory.
However, the working equations can be cast into a very similar form.
The Kohn--Sham eigenvalues are replaced with the so-called quasiparticle energies
from the $GW$ method \cite{Golze.Dvorak.ea:GW.2019} and the electronic Hessian does not
depend on the XC kernel but the BSE kernel which effectively screens the Coulomb interaction
for a given frequency. \cite{Blase.Duchemin.ea:Bethe-Salpeter.2020}
Therefore, the computational protocol and the costs are similar
if suitable approximations such as a static screening are applied to compute the BSE kernel.
Then, the only new step in the workflow is the evaluation of the $GW$ quasiparticle
energies.

The $GW$-BSE has become a major competitor of TD-DFT, with even the static screened
kernel outperforming TD-DFT on many occasions at the same computational cost,
especially when targeting charge-transfer or Rydberg excitations.
\cite{Gui.Holzer.ea:Accuracy.2018, Li.Golze.ea:Combining.2022, Blase.Duchemin.ea:BetheSalpeter.2018}
As charge-transfer processes are rather common in extended systems, the added
reliability of optical excitations obtained from the BSE is highly valued among
researchers.
Unlike for DFT and TD-DFT, which have seen 30 years of research being
dedicated towards accurately predicting various other properties like
non-linear optical properties, direct assessment of excited-state properties
and geometries, or even properties unrelated to optical spectra as, for example,
nuclear magnetic resonance (NMR) properties, the applicability of the BSE to these
properties is still rather narrow.
This review will therefore focus on this extended set of molecular properties,
and outline the development and application of advanced algorithms enabling the
BSE to be applied to a more diverse set of common tasks in computational spectroscopy.
An overview of molecular property operators that have already been assessed within the
framework of the BSE is given in Table~\ref{tab:operators}.

In the subsequent sections, we will refer to operators generally as $\hat{v}$, with
Table~\ref{tab:operators} providing the appropriate notation. Molecular properties
are subsequently obtained from these operators in a linear or on-linear response
framework. This allows to transfer many algorithmic developments for TD-DFT to the
BSE framework, transforming it into a full-fledged computational tool for chemistry, 
physics, and materials science.

\begin{table}[htbp]
    \centering
    \caption{Common property operators for which the BSE has already been applied.
    $\alpha$ is the Sommerfeld ﬁne structure constant, $c$ the speed of light,
    $\vec{r}$ the position vector, $\vec{B}$ the magnetic field, and $\vec{p}$
    the momentum operator.
    $\vec{s}$ denotes the spin of an electron, and $Z_K$ the charge of the
    $K$-th nucleus. SOMF refers to the spin--orbit mean field approach.}
    \label{tab:operators}
    \begin{tabular}{lll}
    \toprule
    Name & Symbol & Operator $\hat{v}$ \\
    \midrule
      electric dipole (length) & $\mu$ & $\vec{r}$  \\
      electric dipole (velocity) & $p$ & $- \text{i} \vec{\nabla}$  \\
      magnetic dipole & $m$ & $\vec{r} \times \vec{B} $ \\
      Fermi contact & $h^{\text{FC}}_K$ & $\left({8 \pi \alpha^2}/{3}\right) \delta(\vec{r}_{K}) \vec{s}$ \\
      spin-dipole & $h^{\text{SD}}_K$ & $\alpha^2 \left({3\vec{r}_{K}^{\dagger} \vec{s} ~ \vec{r}_{K} - {r}_{K}^2 \vec{s}}\right)/{{r}_{K}^5}$ \\
      paramagnetic spin-orbit & $h^{\text{PSO}}_K$ & $- \text{i}\alpha^2 \left({\vec{r}_{K} \times \vec{\nabla}}\right)/{{r}_{K}^3}$ \\
      one-electron SOMF & $h^{\text{SOMF}}_1$ &  $\sum_K Z_K \left( \vec{r}_K \times \vec{p}\right)/ 2c^2 r_K^{\,3}$ \\
      two-electron SOMF & $h^{\text{SOMF}}_2$ & $\left({\vec{r} \times \vec{p}}\right)/2 c^2 {r^{\,3}}$ \\
    \bottomrule
    \end{tabular}
\end{table}

\section{Excited States from the BSE}
\label{sec:exstates}
Before explicitly discussing molecular properties, we first need to have a short look at the 
underlying poles of the linear response problems, with the latter also resembling 
the excited states. Within the $GW$-BSE method, excited states can be extracted from 
the general eigenvalue problem
\cite{Rohlfing.Louie:Excitonic.1998, Faber.Boulanger.ea:Many-body.2013,
Blase.Duchemin.ea:BetheSalpeter.2018, Holzer.Klopper:Ionized.2019} 
\begin{equation}
\label{eq:fullBSE}
    \begin{pmatrix}
    \mathbf{A}^{\phantom{*}}(\Omega) & \mathbf{B}^{\phantom{*}}(\Omega)\\
    \mathbf{B}^*(\Omega) & \mathbf{A}^*(\Omega)
    \end{pmatrix}
    \begin{pmatrix}
        {X} \\
        {Y}
    \end{pmatrix}
    = 
    \Omega 
    \begin{pmatrix}
    \mathbf{1} & \phantom{-}\mathbf{0} \\
    \mathbf{0} & -\mathbf{1} 
    \end{pmatrix}
    \begin{pmatrix}
        {X} \\
        {Y}
    \end{pmatrix}
\end{equation}
with the electronic Hessian consisting of the the matrices $\mathbf{A}$ and $\mathbf{B}$
being defined as\cite{Bintrim.Berkelbach:Full-frequency.2022}
\begin{align}
\label{eq:Afull}
    A_{ai,bj} = & (\epsilon_a - \epsilon_{{i}}) \delta_{ab} \delta_{{i}{j}} + (ai|jb) - (ab|ji) - \Xi_{ab,ji}(\Omega) \\
\label{eq:Bfull}
    B_{ai,bj} = & (ai|bj) - (aj|bi) - \Xi_{bi,aj}(\Omega)
\end{align}
and the solution vectors $\{ \mathbf{X} \mathbf{Y} \}$ being normalized to
the condition
\begin{equation}
    X_N X_M^{\dagger} - Y_N Y_M^{\dagger} = \delta_{NM}
\end{equation}
where $N$ and $M$ denote different solutions or excited states. $\epsilon_i$ and $\epsilon_a$
are the $GW$ quasiparticles of the occupied ($i, j, \dots$) and the virtual
states ($a, b, \dots$).
These may be calculated with the one-shot $G_0W_0$, the iterative
eigenvalue-only self-consistent ev$GW$, or the full quasiparticle self-consistent
qs$GW$ approximation. \cite{Kaplan.Harding.ea:Quasi-Particle.2016}
The frequency-dependent BSE kernel $\boldsymbol{\Xi}(\Omega)$ is given as
\begin{equation}
\label{eq:fullfreq}
    \Xi_{pq,rs} (\Omega) =  \frac{\text{i}}{2\pi} \int \text{d} \omega' e^{- \text{i} \Omega 0^{+}} W_{pq,rs} (\Omega)
                            \left[ \frac{1}{\Omega - \omega' - (\epsilon_q - \epsilon_{{r}}) + \text{i}\eta}
                          + \frac{1}{\Omega + \omega' - (\epsilon_p - \epsilon_{{s}}) + \text{i}\eta} \right]
\end{equation}
where $p,q,r,s$ denote general states.
Here, the screened Coulomb interaction $\mathbf{W}$ is defined as
\begin{equation}
\label{eq:W_BSE}
    W_{pq,rs} (\Omega) = \sum_{tu} \kappa_{pq,tu}^{-1}(\Omega) (tu|rs)
\end{equation}
with the dielectric function $\kappa$. The screened Coulomb interaction is
commonly used also in the preceding $GW$ step within a $GW$-BSE implementation.
Efficient procedures to evaluate Eq.~\ref{eq:W_BSE} for arbitrary values of 
$\Omega$ have been outlined in literature, and especially $\Omega = 0$ can be 
evaluated very efficiently as it is always Hermitian positive definite.
\cite{Holzer.Klopper:Ionized.2019, Holzer:Practical.2023}
We note that the first common approximation has already been applied to 
Eqs.~\ref{eq:fullBSE} to \ref{eq:fullfreq}. To arrive at this result, 
the BSE kernel is approximated as\cite{Blase.Duchemin.ea:Bethe-Salpeter.2020}
\begin{equation}
\label{eq:kernel}
    \text{i} {f}^{\text{BSE}} = \mathbf{v} + \frac{\partial {\Sigma}}{\partial \mathbf{G}} \overset{\text{GWA}}{\longrightarrow} \mathbf{v} + \frac{\partial \mathbf{GW}}{ \partial \mathbf{G}} \overset{\frac{\partial W}{\partial G} = 0}{\longrightarrow} \mathbf{v} - \boldsymbol{\Xi}(\Omega)
\end{equation}
subsequently applying the $GW$ approximation (GWA) to the self-energy $\Sigma$
neglecting the partial derivative resulting from the screened Coulomb 
interaction $\mathbf{W}$ being also implicitly dependent on the Green's function 
$\mathbf{G}$. In Eq.~\ref{eq:kernel}, $\mathbf{v}$ denotes the bare Coulomb interaction, giving rise to an exchange term.
Solving the frequency dependent Eq.~\ref{eq:fullBSE} is still rather involved, 
with the calculation of Eq.~\ref{eq:fullfreq} being the time-limiting step.
Therefore, usually another approximation is invoked, 
\begin{equation}
    \boldsymbol{\Xi}(\Omega) \approx \mathbf{W}(\Omega = 0)
\end{equation}
This leads to the the static screened approximation of the BSE, 
reducing the computational demands by one to two orders of magnitude
if approximations such as the resolution-of-the-identity (RI) approximation
\cite{Ren.Rinke.ea:Resolution-of-identity.2012} are used.
Compared to the fully dynamic BSE method, 
errors of 0.1 to 0.3$\,$eV are observed\cite{Rohlfing.Louie:Electron-hole.2000},
outlining that the applied approximations are not entirely harmless, 
but more of a necessity to make the BSE competitive with 
time-dependent density functional theory.
Further analyzing the matrix elements of Eq.~\ref{eq:Afull} and \ref{eq:Bfull}
in more detail, Rohlfing and Louie noted that the direct part
of the electron-hole interaction kernel involving $\mathbf{W}$
mainly controls the interactions between the quasihole and
quasiparticle states, while the exchange parts with $\mathbf{v}$ tune
the splitting between singlet and triplet excitations.
\cite{Rohlfing.Louie:Electron-hole.2000} In the last few years, 
a notable deficiency of the static screened BSE approximation
to describe triplet states has been noted, and indeed this can
be traced back to an overscreening from the bare Hartree--Fock (HF)
exchange term, arising from neglecting higher-order derivatives
of the screened exchange.
Even given this partly severe drawbacks, the static screened BSE has 
become the de-facto standard in current applications of the $GW$-BSE method in both
theoretical chemistry and solid state physics, which can mostly be attributed to
the superior numerical scaling of its implementations. Current implementation
manage to predict optical spectra sizable systems composed of well over 100 atoms.
\cite{Leng.Jin.ea:GW.2016, Blase.Duchemin.ea:Bethe-Salpeter.2020,
Forster.Visscher:Quasiparticle.2022, Yu.Jin.ea:GPU-Accelerated.2024,
Holzer:Practical.2023}

\section{Simplifications for Real-Valued Orbitals}
\label{sec:real_orbitals}

Similar to TD-DFT, for real-valued molecular orbitals, the generalized BSE
eigenvalue problems can be converted into a set of two symplectic eigenvalue 
problems. This allows for a more concise calculation of properties in 
many cases. The two coupled symplectic eigenvalue problems read
\begin{align}
\label{eq:linear_symplectic+}
\left[ (\mathbf{A} - \mathbf{B}) (\mathbf{A} + \mathbf{B}) \right] {(X+Y)} = & \Omega^2 {(X+Y)} \,\\
\label{eq:linear_symplectic-}
\left[ (\mathbf{A} + \mathbf{B}) (\mathbf{A} - \mathbf{B}) \right] {(X-Y)} = & \Omega^2 {(X-Y)} \,
\end{align}
with the matrices $\mathbf{(A+B)}$ and $\mathbf{(A-B)}$ simply defined
as linear combinations of the previously outlined matrices $\mathbf{A}$ and $\mathbf{B}$
\begin{align}
\label{eq:apb}
(A+B)_{ai,bj} = & (\epsilon_a - \epsilon_{{i}}) \delta_{ab} \delta_{{i}{j}} + 
H^{+}_{ai,bj} (\Omega) \\
\label{eq:amb}
(A-B)_{ai,bj} = & (\epsilon_a - \epsilon_{{i}}) \delta_{ab} \delta_{{i}{j}} + H^{-}_{ai,bj} (\Omega) 
\end{align}
Within Eqs.~\ref{eq:linear_symplectic+} and \ref{eq:linear_symplectic-},
both the right (${X}+{Y}$)  and left (${X}-{Y}$) solutions are normalized to
obey the relation
\begin{equation}
\langle (X+Y)_N | (X-Y)_M \rangle = \delta_{NM}
\end{equation}
The linear combinations of the Coulomb, unscreened exchange, and screened Coulomb terms read
\begin{align}
\label{eq:bse_kernel}
    H^{+,\text{BSE}}_{pq,rs}(\Omega) =&\ (pq|rs) - (ps|rq) - (pr|sq) - \Xi_{ps,rq}(\Omega) - \Xi_{pr,sq}(\Omega) \\
    H^{-,\text{BSE}}_{pq,rs}(\Omega) =&\ (ps|rq) - (pr|sq) + \Xi_{ps,rq}(\Omega) - \Xi_{pr,sq}(\Omega) 
\end{align}
These linear combinations provide the significant
advantage of being either symmetric ($+$) or skew-symmetric ($-$) with respect to
interchanges in indices. They can therefore be directly related to properties
with symmetric (real) operators, as for example electric fields, or to
properties with skew-symmetric (imaginary) operators, as for example
magnetic fields. While rewriting the eigenvalue problem is also possible for complex
molecular orbitals, this symmetry correspondence is lost. Therefore, 
in the case of complex orbitals, the advantages of using Eqs.~\ref{eq:linear_symplectic+}
and \ref{eq:linear_symplectic-} over Eq.~\ref{eq:fullBSE} are limited. 
\cite{Holzer.Pausch.ea:GWBSE.2021}
Complex orbitals are needed for, e.g., calculations in finite magnetic fields
or relativistic approaches treating spin--orbit coupling variationally in the
ground state.

\section{Optical Linear Response Properties}
\label{sec:linear_properties}

In linear response theory, the corresponding linear response function
of a system with respect to a time-dependent field oscillating with the 
frequency $\omega$ can be written as
\cite{Ghosh.Deb:Dynamic.1982, Jamorski.Casida.ea:Dynamic-polarizabilities.1996}
\begin{equation}
    \langle \langle v^{\zeta}; v^{\eta}(\omega) \rangle \rangle = 
    \Tr(\hat{v}^{\zeta} \gamma^{\eta}(\omega))
\end{equation}
with the property operator $\hat{v}^{\zeta}$ describing the perturbation
$\zeta$ and the first-order reduced density matrix $\gamma^{\eta}$ of the perturbation $\eta$,
\begin{equation}
    \gamma^{\eta}  = 
    \begin{pmatrix}
    0 & X^{\eta} \\
    Y^{\eta} & 0
    \end{pmatrix}
\end{equation}
Using these definitions, determining linear optical response properties 
from the BSE is straightforward, as one simply needs to modify the general 
eigenvalue problem to instead determine the components of the 
frequency-dependent first-order transition density matrices. Assuming that the general 
property vectors $P$ and $Q$ collect the integrals of the external perturbation, 
\begin{equation}
\label{eq:propints}
    \langle \phi_a | \hat{v}^{\eta}| \phi_i \rangle =  P_{ai} = Q_{ia}^*
\end{equation}
the coupled-perturbed BSE equations reads
\begin{equation}
\label{eq:CPBSE}
    \left[
    \begin{pmatrix}
    \mathbf{A}^{\phantom{*}}(\omega) & \mathbf{B}^{\phantom{*}}(\omega)\\
    \mathbf{B}^*(\omega) & \mathbf{A}^*(\omega)
    \end{pmatrix}
    -\omega 
    \begin{pmatrix}
    \mathbf{1} & \phantom{-}\mathbf{0} \\
    \mathbf{0} & -\mathbf{1} 
    \end{pmatrix}
    \right]
    \begin{pmatrix}
        {X} \\
        {Y}
    \end{pmatrix}
    = 
    \begin{pmatrix}
        {P} \\
        {Q}
    \end{pmatrix}
\end{equation}
The right-hand side (RHS) of Eq.~\ref{eq:CPBSE} is solely determined by the 
external perturbation, e.g. an electric field oscillating with the frequency
$\omega$ and not by the excited states $\Omega$. In Ref.~\citenum{Kehry.Franzke.ea:Quasirelativistic.2020}, 
it was therefore outlined that in cases where many excited states exist in or before the energetic 
region of interest, Eq.~\ref{eq:CPBSE} can be used to efficiently bypass the overhead
of calculating many excited states from the BSE eigenvalue problem. To obtain an
optical spectrum, consequently the frequency-dependent dipole polarizability
is calculated directly from the solutions of Eq.~\ref{eq:CPBSE} via a
direct product,
\begin{equation}
\label{eq:polarizability}
    \alpha^{\eta \zeta} (\omega) = \langle P^{\eta}, Q^{\eta} | X^{\zeta}(\omega), Y^{\zeta}(\omega) \rangle
\end{equation}
Within the full frequency-dependent BSE, this is even simpler than
the general eigenvalue problem: Unlike the energies of the excited states $\Omega$, the frequency
of the external field $\omega$ is already known \textit{a priori}! 
Therefore, the left-hand side (LHS) of the coupled-perturbed equation \ref{eq:CPBSE}
does not change during an iterative procedure.
If real orbitals are used as outlined in the previous section, the coupled-perturbed
linear response equation can be rewritten as
\begin{align}
\label{eq:cpbse_symplectic1}
\left[ (\mathbf{A}(\omega) + \mathbf{B}(\omega)) (X+Y) - \omega {(X-Y)} \right] = & -(P+Q) \,  \\
\label{eq:cpbse_symplectic2}
\left[ (\mathbf{A}(\omega) - \mathbf{B}(\omega)) (X-Y) - \omega {(X+Y)} \right] = & -(P-Q) \, 
\end{align}
From Eqs.~\ref{eq:propints}, \ref{eq:cpbse_symplectic1}, and \ref{eq:cpbse_symplectic2}
it can be concluded that $(P+Q)$ is zero for purely imaginary operators, while $(P-Q)$ is zero 
for purely real operators. Note that for any non-vanishing frequency $\omega$, 
both equations are strictly coupled and neither part can be neglected. Still, 
instead of one eigenvalue problem of full size, two with half the original 
size can be solved. Caution must, however, be used, as in the limiting case of the 
frequency of the external perturbation approaching an excited state, 
$\omega \rightarrow \Omega$, the coupled-perturbed
Bethe--Salpeter equations will diverge. It is therefore often necessary to
assume that the external perturbation has a constant imaginary component, 
$\omega = \omega_r + \Gamma$, ``damping'' the diverging cases. The assumed imaginary 
component directly leads to a broadening of the polarizability spectra 
with a Lorentzian line shape.

\begin{figure}[htbp]
  \includegraphics[width=\columnwidth]{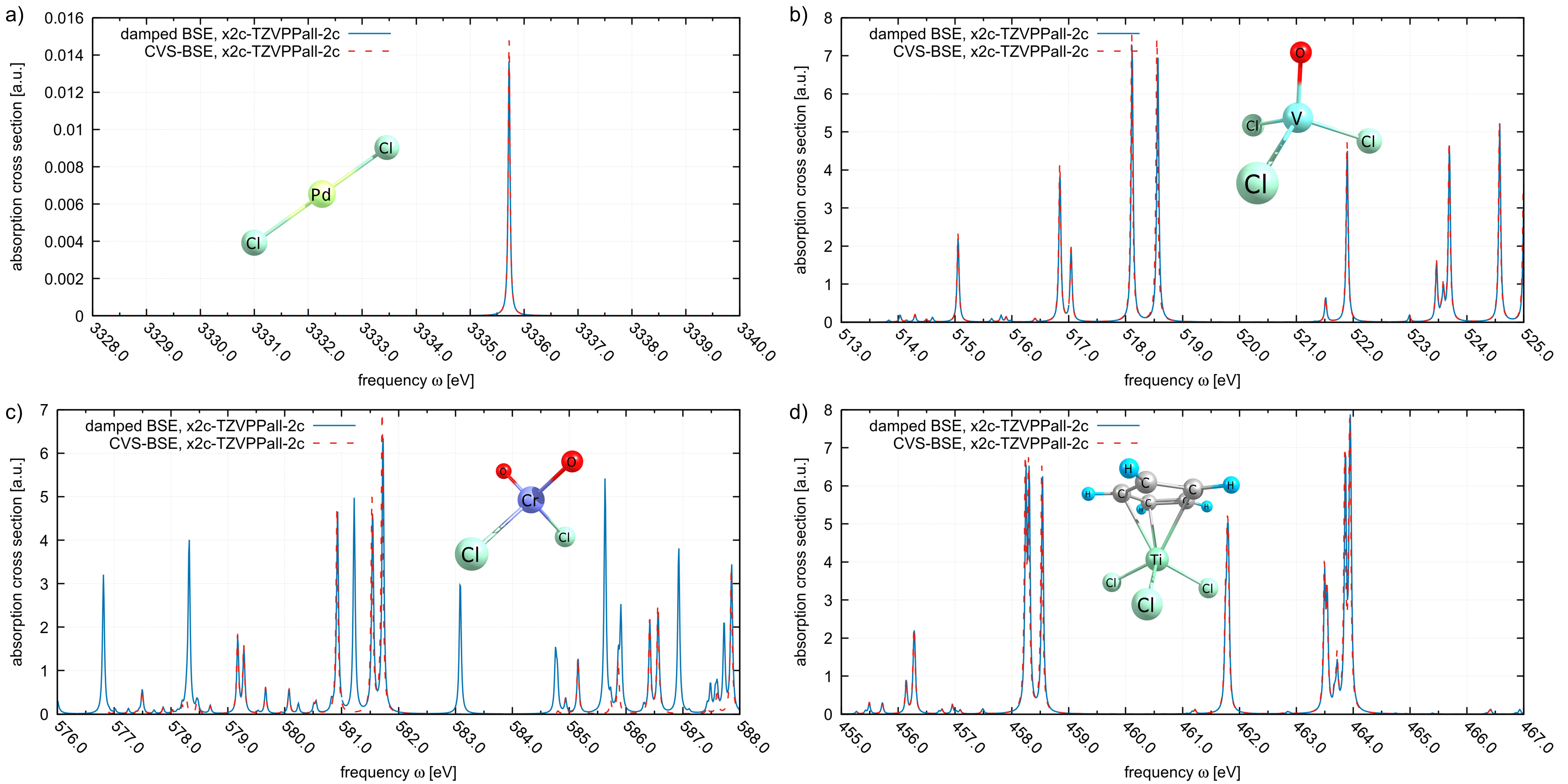}
  \caption{Damped response BSE (blue solid lines) and CVS-BSE XAS spectra (red dashed lines) comparing the
             energy-integrated absorption cross section of a) the PdCl$_2$ L$_2$-edge,
             b) the VOCl$_3$ L$_{2,3}$-edges, 
             c) the CrO$_2$Cl$_2$ L$_{2,3}$-edges, and d) the TiCpCl$_3$ L$_{2,3}$-edges.
             All spectra calculated at the frequency-sampled contour deformation
             $G_0W_0${\makeatletter @}TMHF level of theory. A damping parameter of 0.02~eV was 
             used for damped response BSE. CVS-BSE spectra were broadened accordingly with Lorentzians.
             Absorption cross section given in atomic units (= Hartree $\cdot$ bohrs$^2$), 
             frequency in eV. Scalar-relativistic and spin--orbit effects are introduced
             with exact two-component (X2C) theory.
             Reprinted from M.\ Kehry, W.\ Klopper, C.\ Holzer,
             \textit{J.\ Chem.\ Phys.}\ \textbf{2023}, \textit{159}, 044116.
             Copyright the Authors. Licensed under a Creative Commons Attribution (CC BY) 
             license (\url{https://creativecommons.org/licenses/by/4.0/)}.}
  \label{fig:L-edges}
\end{figure}

The ability to bypass lower-lying states can for example 
be utilized to assess core excited states without neglecting or projecting out valence 
orbitals, as shown in Figure~\ref{fig:L-edges}. Given the high performance of the
$GW$-BSE  method for core excited states,\cite{Kehry.Klopper.ea:Robust.2023}
this is a significant progress in the prediction of K- or L-edges.
Figure~\ref{fig:L-edges} further outlines that indeed common approximations
used, such as the core-valence separation (CVS) approximation, are usually
well suited for the extraction of core excited states, though the
implementation of complex dynamic polarizabilities allows for a
convenient yet efficient way of checking this approximation if there 
is doubt about its reliability.

The equations outlined in this section can generally be used to calculate 
linear response properties of molecular systems, as long as one can formulate the
appropriate RHS in terms of integrals also expressible in the same basis
of molecular orbitals. This encompasses, for example, static and dynamic
polarizabilities, optical rotation tensors, or dynamic magnetizabilities.
To illustrate this point, results for static dipole polarizabilities of the metallocenes
FeCp$_2$, RuCp$_2$, and OsCp$_2$ are presented in Table~\ref{tab:polarizabilities}.
Here, the electric dipole operator represents the perturbations $\zeta$ and $\eta$.
The Kohn--Sham DFT approach underestimates the polarizabilities of all three complexes
and the $GW$-BSE reduces the deviation towards the experiment. \cite{Goebel.Hohm:Comparative.1997}
$GW$-BSE{\makeatletter @}PBE is able to recapture the experimental trend of
a monotonous increase of the polarizability with increasing mass for these
metallocenes. The good performance of $GW$-BSE is not restricted to static
polarizabilities but also dynamic polarizabilities are obtained in good
agreement with the experiment. \cite{Kehry.Franzke.ea:Quasirelativistic.2020,
Rocca.Lu.ea:Ab-initio-calculations.2010}

\begin{table*}[htbp]
\caption{Static polarizabilities for group 8 metallocenes in atomic units
at the Kohn--Sham DFT and $GW$-BSE levels with the dhf-TZVP basis set.
Computational results are taken from Ref.~\citenum{Kehry.Franzke.ea:Quasirelativistic.2020}
and compared to the experimental findings (Expt.) of Ref.~\citenum{Goebel.Hohm:Comparative.1997}.
}
\label{tab:polarizabilities}
\begin{tabular}{@{\extracolsep{6pt}}
l
S[table-format = 2.1]
S[table-format = 2.1]
S[table-format = 2.1]
S[table-format = 2.1]
S[table-format = 2.1]
c
@{}}
\toprule 
& \multicolumn{2}{c}{{\text{CAM-B3LYP}}} & \multicolumn{2}{c}{{\text{PBE0}}} & \\
\cmidrule{2-3} \cmidrule{4-5}
 & {\text{DFT}} & {\text{$GW$-BSE}} & {\text{DFT}} & {\text{$GW$-BSE}} & {\text{Expt.}} \\
\midrule
FeCp$_2$ & 118.3 & 128.3 & 118.8 & 131.3 & 126.1 \\
RuCp$_2$ & 127.6 & 132.0 & 128.1 & 134.2 & 133.1 \\
OsCp$_2$ & 128.0 & 130.6 & 128.7 & 137.9 & 138.5 \\
\bottomrule
\end{tabular}
\end{table*}

\section{Optical Non-Linear Response Properties}
\label{sec:nonlinear_properties}

Non-linear optical properties are more tedious to obtain. For example, 
for non-linear hyperpolarizabilities, multiple external fields 
with different frequencies $\omega$ act on the system, leading to a
distinctly more involved response of the latter. Accordingly, 
the second-order reduced density matrix
\cite{Parker.Rappoport.ea:Quadratic.2018, Rauwolf.Klopper.ea:Non-linear.2024}
\begin{equation}
    \gamma^{\zeta \eta}  = 
    \begin{pmatrix}
    K^{\zeta \eta} & X^{\zeta \eta} \\
    Y^{\zeta \eta} & K^{\zeta \eta}
    \end{pmatrix}
\end{equation}
is needed to evaluate the quadratic response functions
\begin{equation}
\label{eq:hyperpol}
    \langle \langle v^{\zeta}; v^{\eta}(\omega), v^{\theta}(\omega') \rangle \rangle = 
    \Tr(\hat{v}^{\zeta} \gamma^{\eta \theta}(\omega,\omega'))
\end{equation}
Note that the response is assumed to be instantaneous and independent, 
so that the order in which the frequency-dependent external 
perturbations $v^{\eta}(\omega)$ and $v^{\theta}(\omega')$ are applied does 
not matter. The diagonal matrix elements $\mathbf{K}$ can be obtained
directly from the linear response functions, 
\begin{align}
\label{eq:Kij}
 K_{ij}^{\zeta \eta} = - &\sum_a \left[ X_{aj}^{\zeta} Y_{ai}^{\eta} +  Y_{ai}^{\zeta} X_{aj}^{\eta} \right] \\
\label{eq:Kab}
 K_{ab}^{\zeta \eta} = \phantom{-} &\sum_i \left[ Y_{bi}^{\zeta} X_{ai}^{\eta} +  X_{ai}^{\zeta} Y_{bi}^{\eta} \right]
\end{align}
Contrary, the off-diagonal matrix elements $X^{\zeta \eta}$ and $Y^{\zeta \eta}$
need to be determined from another linear problem of the same structure as 
Eq.~\ref{eq:CPBSE}, with $\omega = \omega^{\zeta} + \omega^{\eta}$
and $P$ and $Q$ replaced by their quadratic response analogues\cite{Rauwolf.Klopper.ea:Non-linear.2024}
\begin{align}
\label{eq:Pia}
    P_{ai}^{\zeta \eta} = & \sum_b \left( U_{ab}^{\zeta} X_{bi}^{\eta} + U_{ab}^{\eta} X_{bi}^{\zeta}  \right)
   -  \sum_j \left( U_{ij}^{\zeta} X_{aj}^{\eta} + U_{ij}^{\eta} X_{aj}^{\zeta}  \right) + g_{ai}^{\text{BSE}}(X^\zeta, Y^{\zeta}; X^{\eta}, Y^{\eta}) \\
\label{eq:Qia}
    Q_{ai}^{\zeta \eta} = & \sum_b \left( V_{ab}^{\zeta} Y_{bi}^{\eta} + V_{ab}^{\eta} Y_{bi}^{\zeta}  \right) 
   -  \sum_j \left( V_{ij}^{\zeta} Y_{aj}^{\eta} + V_{ij}^{\eta} Y_{aj}^{\zeta}  \right) + g_{ia}^{\text{BSE}}(X^\zeta, Y^{\zeta}; X^{\eta}, Y^{\eta})
\end{align}
with
\begin{align}
    U_{pq}^{\zeta} = \sum_{bj} A'_{pq,bj} X_{bj}^{\zeta} + B_{pq,bj} Y_{bj}^{\zeta} + v_{pq}^{\zeta}\\ 
    V_{pq}^{\zeta} = \sum_{bj} B^{*}_{pq,bj} X_{bj}^{\zeta} + A'^{*}_{pq,bj} Y_{bj}^{\zeta} + v_{pq}^{\zeta,*} 
\end{align}
The matrix $\mathbf{A'}$ denotes a slightly modified Eq.~\ref{eq:Afull}, where
the quasiparticle energy differences have been dropped:
\begin{equation}
    A'_{pq,rs} =  (pq|sr) - (pr|sq) - \Xi_{pr,sq}(\Omega)
\end{equation}
Upon inspection of Eqs.~\ref{eq:Kij} to \ref{eq:Qia}, the symmetry upon
switching the perturbations $\zeta$ and $\eta$ is obvious, again outlining
that the order of applying the perturbations does not matter for 
anything other than indexing purposes. We remind the reader that 
again simplifications can be made for purely real orbitals, though it
is not necessary to repeat those, as a detailed outline of the
corresponding formulas has been given in
Ref.~\citenum{Rauwolf.Klopper.ea:Non-linear.2024}.
Another key approximation has been made in the determination
of Eqs.~\ref{eq:Pia} and \ref{eq:Qia} by again neglecting
hyperkernel derivatives. Following the approximations made for the
BSE kernel, which explicitly neglect the derivative of the
screened Coulomb part with respect to the Greens function leads to
\cite{Villalobos-Castro.Knysh.ea:Lagrangian-Z-vector.2023,
Himmelsbach.Holzer:Excited.2024}
\begin{equation}
\label{eq:hyperkernel}
    {g}^{\text{BSE}} = \frac{\partial^2 {\Sigma}}{\partial \mathbf{G} \partial \mathbf{G}} \overset{Eq.\ref{eq:kernel}}{\longrightarrow}  \frac{\partial \mathbf{W}}{ \partial \mathbf{G}} \overset{\frac{\partial W}{\partial G} = 0}{\longrightarrow} 0
\end{equation}
The BSE hyperkernel ${g}^{\text{BSE}}$ is therefore set to zero, though the impact
of this approximation has not yet been assessed in detail.

Table~\ref{tab:hyperpolarizabilities} illustrates the application of the
$GW$-BSE to the hyperpolarizabilities of small molecules. Conventional
DFT methods such as CAM-B3LYP and PBE0 lead to a significant deviation
from the experiment. $GW$-BSE substantially improves the results and
leads to an excellent agreement with the experimental findings when
using well suited Kohn--Sham starting points such as the TMHF
functional, which was also successfully applied in the previous
section for linear response properties. That is, the behavior
in terms of the Kohn--Sham reference is well transferred to higher-order
derivatives, suggesting that a single reference can be used for a wide
range of optical properties in the $GW$-BSE. Based on the given results,
neglecting the BSE hyperkernel also seems to be well justified for
hyperpolarizabilities.

\begin{table*}[t!]
\caption{Dynamic first hyperpolarizability $\beta_{\parallel}$ (in atomic units)
for H$_2$O, MeOH (Me = CH$_3$), and dimethyl ether (DME) at 1064\,nm
calculated at the Kohn--Sham DFT and $GW$-BSE levels with the d-aug-cc-pVQZ basis set.
Computational results are taken from Ref.~\citenum{Rauwolf.Klopper.ea:Non-linear.2024}
and compared to the experimental findings (Expt.) of
Refs.~\citenum{Kaatz.Donley.ea:A-comparison.1998} (H$_2$O, MeOH) and 
\citenum{Couling.Shelton:Hyperpolarizability.2015} (DME).
}
\label{tab:hyperpolarizabilities}
\begin{tabular}{@{\extracolsep{6pt}}
l
S[table-format = -2.1]
S[table-format = -2.1]
S[table-format = -3.1]
c
@{}}
\toprule 
Method & {\text{H$_2$O}} & {\text{MeOH}} & {\text{DME}} & \\
\midrule
CAM-B3LYP & -17.7 & -35.0 & -102.9 \\
PBE0      & -19.5 & -40.7 & -119.3 \\
$G_0W_0$-BSE{\makeatletter @}PBE0 & -20.6 & -42.8 & -126.3 \\
ev$GW$-BSE{\makeatletter @}PBE0   & -18.2 & -38.7 & -111.5 \\
$G_0W_0$-BSE{\makeatletter @}TMHF & -18.6 & -34.8 & -105.0 \\
ev$GW$-BSE{\makeatletter @}TMHF   & -16.7 & -32.5 & -96.4 \\
\midrule
Expt.     & {\text{$-19.2 \pm 0.9$}} & {\text{$-31.2 \pm 1.6 $}} & {\text{$-94.0 \pm 0.25$}} \\
\bottomrule
\end{tabular}
\end{table*}

Besides the hyperkernel, two-photon absorption processes can
also be described within a non-linear regime. Therefore, the
$GW$-BSE is now able to account for all common sorts of excitations.

\section{Transition Moments Between Excited States}
\label{sec:es_properties}

Again similar to TD-DFT, also properties between excited states
are of growing interest, leading to the possibility to predict,
for example, transient absorption. Intrinsically, calculating
a property that involves an expectation value between two
excited states is very similar to calculating non-linear
properties as outlined in the previous section.
The second-order reduced density matrix only needs to be
slightly modified, yielding\cite{Himmelsbach.Holzer:Excited.2024}
\begin{equation}
\label{eq:gammaNM}
    \boldsymbol{\gamma}^{NM}  = 
    \begin{pmatrix}
    \mathbf{K}^{NM} & \mathbf{X}^{NM} \\
    \mathbf{Y}^{NM} & \mathbf{K}^{NM}
    \end{pmatrix}
\end{equation}
From the second-order response matrix, the corresponding
transition property between two excited states $N$ and $M$ 
can be calculated similar to Eq.~\ref{eq:hyperpol}, by choosing to let the frequency of 
the perturbation approach the energy of the excited states, 
\begin{equation}
\label{eq:hyp_trace}
\begin{split}
     v_{NM} =  \lim_{\omega \rightarrow \Omega_N}  (\omega - \Omega_N) \lim_{\omega' \rightarrow -\Omega_M} (\omega' + \Omega_M)
  \times  \langle \langle v^{\zeta}; v^{\eta} (\omega), v^{\theta}(\omega') \rangle \rangle  
  =   \Tr\left( \hat{v}^{\zeta} \gamma^{NM} \right) 
\end{split}
\end{equation}
Note that in Eq.~\ref{eq:hyp_trace} the selection of excited states
matter, and generally $\boldsymbol{\gamma}_{NM} \ne \boldsymbol{\gamma}_{MN}$. 
The matrix elements of $\boldsymbol{\gamma}_{NM}$ can be evaluated similar
to the previous section with the matrix elements of $\mathbf{K}$ being defined as
\begin{align}
\label{eq:KijNM}
 K_{ij}^{NM} = - &\sum_a \left[ X_{ja}^{N} X_{ai}^{*M} + Y_{ai}^{N} Y_{aj}^{*M} \right] \\
\label{eq:KabNM}
 K_{ab}^{NM} = \phantom{-} &\sum_i \left[ X_{bi}^{N} X_{ai}^{*M} + Y_{ai}^{N} Y_{bi}^{*M} \right] 
\end{align}
The corresponding RHS needed to evaluate Eq.~\ref{eq:CPBSE} is then obtained as
\begin{align}
\label{eq:PiaNM}
    P_{ai}^{NM} = & \sum_b \left( U_{ab}^{N} X_{bi}^{*M} + U_{ab}^{M} Y_{bi}^{N}  \right)
   -  \sum_j \left( U_{ij}^{N} X_{aj}^{*M} + U_{ij}^{M} Y_{aj}^{N}  \right) \\
\label{eq:QiaNM}
    Q_{ai}^{NM} = & \sum_b \left( V_{ab}^{N} Y_{bi}^{*M} + V_{ab}^{M} X_{bi}^{N}  \right) 
   -  \sum_j \left( V_{ij}^{N} Y_{aj}^{*M} + V_{ij}^{M} X_{aj}^{N}  \right)
\end{align}
with
\begin{align}
    U_{pq}^{N} = & \sum_{bj} A'_{pq,bj} X_{bj}^{N} + B_{pq,bj} Y_{bj}^{N}\\ 
    V_{pq}^{N} = & \sum_{bj} B^{*}_{pq,bj} X_{bj}^{N} + A'^{*}_{pq,bj} Y_{bj}^{N}
\end{align}
The hyperkernel contribution has again been dropped. 
The frequency of at which Eq.~\ref{eq:CPBSE} needs to be 
evaluated with the constructed RHS is exactly 
$\omega = \Omega_M - \Omega_N$.
If the excited states $N$ and $M$ are switched in Eq.~\ref{eq:hyp_trace}, 
instead of the eigenpair $\omega, \{\mathbf{X}$, $\mathbf{Y} \}$, 
the time-reversal symmetry related eigenpair 
$-\omega, \{\mathbf{Y}^{*}$, $\mathbf{X}^{*} \}$ 
is to be used. As a consequence, upon the interchange $N \leftrightarrow M$, 
also $\mathbf{X} \rightarrow \mathbf{Y}^{\dagger}$ and $\mathbf{Y} \rightarrow \mathbf{X}^{\dagger}$
must be interchanged accordingly in the second-order reduced density matrix.
Evaluating the matrices $\mathbf{K}$ with interchanged indices 
also leads to the corresponding adjoint matrix, resulting in the relation 
$\boldsymbol{\gamma}_{NM} = (\boldsymbol{\gamma}_{MN})^{\dagger}$.
Transition properties originating from purely real operators are
therefore unaffected by interchanging excited states $N$ and $M$, while 
purely imaginary transition properties switch signs.

\begin{figure}[ht!]
  \includegraphics[width=0.75\columnwidth]{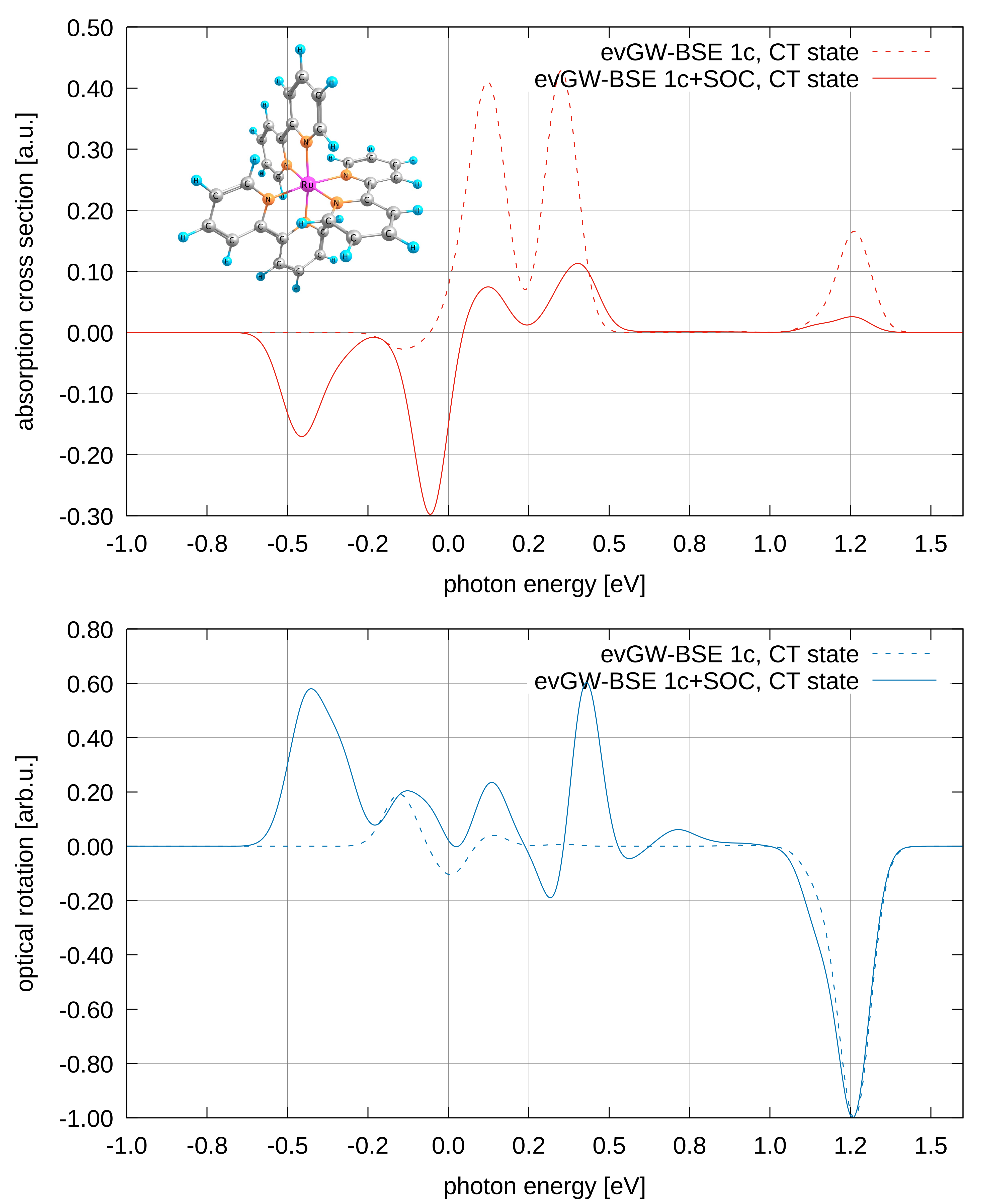}
  \caption{Predicted ev$GW$--BSE transient absorption (upper) 
    and optical rotation (lower) spectra of 
    [Ru(bpy)$_3$]$^{2+}$
    obtained from scalar-relativistic (1c, dotted line) and 
    scalar-relativistic plus perturbative spin--orbit coupling 
    (1c+SOC, solid line). Negative oscillator
    strengths correspond to emission lines.
    An arbitrary broadening of 0.05$\,$eV is applied.
    The absorption cross section at 0.0$\,$eV is an 
    artifact of this broadening.
    Absorption cross section given in atomic units (a.u.), 
    optical rotation in arbitrary units (arb.u.).
    Reprinted (adapted) from P.\ Himmelsbach, C.\ Holzer,
    \textit{J.\ Chem.\ Phys.}\ \textbf{2024}, \textit{161}, 241105.
    Copyright the Authors. Licensed under a Creative Commons Attribution
    NonCommercial 4.0 International (CC BY-NC) license 
    (\url{https://creativecommons.org/licenses/by-nc/4.0/)}.}
  \label{fig:Rubpy}
\end{figure}

Calculating transition moments between excited states is of 
high interest in the field of optical materials, as for 
example, in the design of organic light emitting diodes (OLEDs).
The prototype complex tris-bipyrdin ruthenium [Ru(bpy)$_3$]$^{2+}$
acts via an intersystem crossing (ISC) mechanism
starting from an excited ligand-to-metal charge transfer (LMCT)
state, which is hard to capture correctly with TD-DFT unless
careful functional tuning and testing is carried out.
$GW$-BSE on the other hand can perfectly reproduce
these LMCT states, and also yield sufficient predictive
power for excited-state transitions, even including
ISC when spin--orbit coupling is included. Figure~\ref{fig:Rubpy}
outlines the transient absorption spectra simulated using
ev$GW$-BSE, with and without including spin--orbit coupling (SOC).
Clearly, including spin--orbit coupling is instrumental
in finding the correct S$\rightarrow$T transition.
Having a method at hand that is able to describe both
LMCT transitions and SOC effects is highly valuable
for researching advanced materials for light-matter interactions, 
and the $GW$-BSE method is currently progressing
to the forefront of this branch of research.

\section{Excited-State Dipole Moments and an Outlook on Gradients}
\label{sec:es_gradients}

\begin{figure}[ht!]
  \includegraphics[width=\columnwidth]{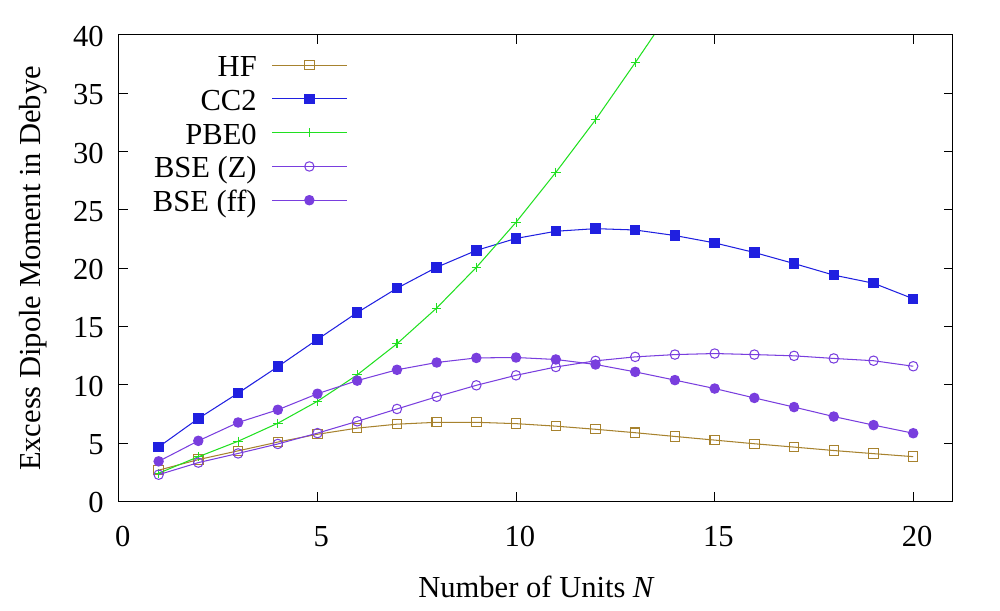}
  \caption{Excess dipole moment of the S$_1$ excited state of the
  push-pull oligomers $\text{H}_2\text{N}-\left[\text{CH}=\text{CH}\right]_N-\text{NO}_2$
  for various system sizes. Calculations are performed at the time-dependent HF,
  linear-response CC2 (relaxed), PBE0 TDDFT, and the ev$GW$-BSE{\makeatletter @}PBE0
  levels using the cc-pVTZ basis set. BSE results are shown for the finite-field (ff)
  approach using a 5-point stencil formula and the analytical Z-vector (Z) approach.
  The excess dipole moment is defined as the norm of the vector difference in
  Eq.~\ref{eq:excited-dipole}. Individual data for plot taken from
  Refs.~\citenum{Villalobos-Castro.Knysh.ea:Lagrangian-Z-vector.2023} (BSE Z-vector)
  and \citenum{Knysh.Villalobos-Castro.ea:Exploring.2023} (other methods).
  }
  \label{fig:excited-dipoles}
\end{figure}

Recently, Ref.~\citenum{Villalobos-Castro.Knysh.ea:Lagrangian-Z-vector.2023} 
has devised a way to also calculate excited-state 
dipole moments, which are conceptually very close also to
analytic excited state gradients. The excited-state dipole moments
can be assessed in an indirect way, calculating the difference
of the excited-state dipole moment and ground-state dipole moment, 
\begin{equation}
\label{eq:excited-dipole}
 \mu_{\alpha}^{N} - \mu_{\alpha}^{0}  = \frac{\partial \Omega^N}{\partial E_{\alpha}} 
\end{equation}
with the Cartesian component $\alpha$ and the excited state $N$.
The difference dipole moment can be determined from the
derivative of the excited-state energy with respect to an
electric field $E_{\alpha}$. While Villalobos-Castro \textit{et al.}
have used a Lagrangian Z-vector formalism in 
Ref.~\citenum{Villalobos-Castro.Knysh.ea:Lagrangian-Z-vector.2023}
to arrive at an explicit formulation, the coupled-perturbed
approach outlined for non-linear properties
can also be used. Assuming that the transition of interest is
the $N \leftrightarrow N$ transition, the excited-state dipole moment
is obtained as
\begin{equation}
\label{eq:exdip}
     \mu^{N}_{\alpha} = \text{tr}\left( \hat{\mu}_{\alpha} \bar{\gamma}^{NN} \right)
\end{equation}
Following Ref.~\citenum{Villalobos-Castro.Knysh.ea:Lagrangian-Z-vector.2023}, 
the corresponding coupled-perturbed equation that needs to be solved is
similar to those needed to be solved for excited-state properties.
$\bar{\gamma}^{NN}$ differs from $\gamma^{NN}$ just by the modification
of the RHS in Eqs.~\ref{eq:PiaNM} and \ref{eq:QiaNM}, 
now including the contribution of the ground-state Hamiltonian, yielding
\begin{align}
\allowdisplaybreaks
\label{eq:PiaNM_Grad}
    P_{ai}^{NM} = & \sum_b \left( U_{ab}^{N} X_{bi}^{*M} + U_{ab}^{M} Y_{bi}^{N}  \right)
   -  \sum_j \left( U_{ij}^{N} X_{aj}^{*M} + U_{ij}^{M} Y_{aj}^{N}  \right) + \sum_{bj} H^{\text{KS}}_{ai, jb} K_{bj} \\
\label{eq:QiaNM_Grad}
    Q_{ai}^{NM} = & \sum_b \left( V_{ab}^{N} Y_{bi}^{*M} + V_{ab}^{M} X_{bi}^{N}  \right) 
   -  \sum_j \left( V_{ij}^{N} Y_{aj}^{*M} + V_{ij}^{M} X_{aj}^{N} \right) + \sum_{bj} H^{\text{KS}}_{ai, bj} K_{bj}
\end{align}
Contributions from the quasiparticle $GW$ step to the RHS have been neglected.
\cite{Villalobos-Castro.Knysh.ea:Lagrangian-Z-vector.2023}
$\mathbf{H}^{\text{KS}}$ denotes the Kohn--Sham kernel of the underlying
density functional approximation.\cite{Bauernschmitt.Ahlrichs:Stability.1996}
The resulting solution vector $\{ \mathbf{X}, \mathbf{Y} \}$ is often referred
to as Z-vector in gradient calculations, and can be used to determine both
excited-state dipole moments as well as analytic gradients, though
for the latter some more ingredients are needed.
\cite{Villalobos-Castro.Knysh.ea:Lagrangian-Z-vector.2023}
At the time of writing, the Z-vector equations have only seen exploratory use
to determine excited-state dipole moments with great success.
\cite{Knysh.Villalobos-Castro.ea:Excess.2023, Knysh.Villalobos-Castro.ea:Exploring.2023}
This is illustrated for the excess dipole moment (S$_1$ excited state) of 
push-pull oligomers in Figure~\ref{fig:excited-dipoles}.
The comparison between the finite-field (ff) calculations and the
Z-vector (Z) approach outlines that the latter leads to the qualitatively
correct behavior, however, setting the screened Coulomb potential to its
zero-field value and approximating the ev$GW$ quasiparticle energies with
the Kohn--Sham values leads to notable deviations. Compared to the underlying
Kohn--Sham methods, which diverges, this Z-vector ansatz still leads to a
tremendous improvement, as the correct behavior of the excess dipole moment
having a maximum at a certain chain length is recovered. PBE0 fails to recover
this trend completely, leading to  unphysical increases of the excess dipole
moment with increased chain length.
This divergence is due to the too small amount of exact exchange in the
long-range region and could be removed with range-separated or optimally
tuned functionals. However, the $GW$-BSE naturally resolves all issues in
a more rigorous way from first principles.

The explicit usefulness of gradients remains to be shown, but initial
investigations point at BSE gradients being very useful, 
\cite{Caylak.Baumeier:Excited-State.2021, Ismail-Beigi.Louie:Excited-State.2003}
though the necessary $GW$ contribution to the gradient remains problematic to date.
\cite{Villalobos-Castro.Knysh.ea:Lagrangian-Z-vector.2023}

\section{NMR Spin--Spin Coupling Constants}
\label{sec:nmr_properties}
So far, this mini-review exclusively presented the application of the
$GW$-BSE to optical properties or light-matter interactions.
Given the success in this field, applications to other research areas
are of interest. NMR spin--spin coupling constants (SSCCs) are closely
related to optical excitations in terms of a computer implementation. 
\cite{Helgaker.Jaszunski.ea:Ab.1999}
The total SSCC consists of three first-order response terms, namely
the Fermi-contact (FC), spin-dipole (SD), and the paramagnetic spin-orbit
(PSO) term, as well as the diamagnetic spin-orbit (DSO) interaction,
which is directly available from the ground-state density matrix.
Usually, the latter is small and the coupling constant is dominated
by the response terms. Here, the FC and SD interactions act as a
real triplet operator, while the PSO term is of imaginary singlet
character. This allows to apply the $GW$-BSE in a post-Kohn--Sham
fashion to improve the DFT performance for these three terms
by solving the static linear response equation
\cite{Franzke.Holzer.ea:NMR.2022}
\begin{equation}
\label{eq:static-CPBSE}
    \begin{pmatrix}
    \mathbf{A}^{\phantom{*}} & \mathbf{B}^{\phantom{*}}\\
    \mathbf{B}^*& \mathbf{A}^*
    \end{pmatrix}
    \begin{pmatrix}
        {X} \\
        {Y}
    \end{pmatrix}
    = 
    \begin{pmatrix}
        {P} \\
        {Q}
    \end{pmatrix}
\end{equation}
where the RHS $P_{ai} = Q_{ia}^*$ denotes the respective perturbation operator
(FC, SD, PSO) in the molecular orbital basis, c.f.\ Eq.~\ref{eq:propints}.
The NMR coupling tensor is then obtained by the static form of Eq.~\ref{eq:polarizability}.
Just like the dynamic generalization in Eq.~\ref{eq:CPBSE}, this equation
can be simplified again and the perturbations operators can be treated separately
for real-valued orbitals, i.e.\ the FC and SD terms are available from the symmetric
linear combination and the PSO interaction is obtained from the skew-symmetric one.
This way, the NMR coupling constants are efficiently obtained in a non-relativistic 
\cite{Helgaker.Jaszunski.ea:Ab.1999} or scalar-relativistic framework. 
\cite{Franzke:Reducing.2023}
For complex-valued orbitals, which are obtained in self-consistent relativistic
spin--orbit calculations, \cite{Franzke.Mack.ea:NMR.2021} the terms are coupled
and Eq.~\ref{eq:static-CPBSE} is solved without simplifications.

\begin{figure*}[htbp]
    \includegraphics[width=0.49\columnwidth]{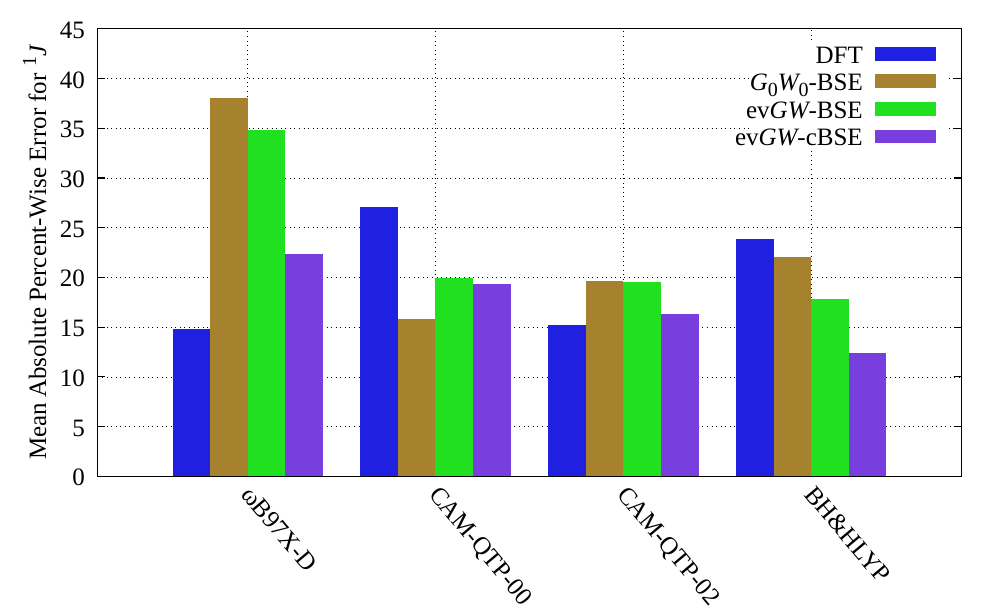}
    \includegraphics[width=0.49\columnwidth]{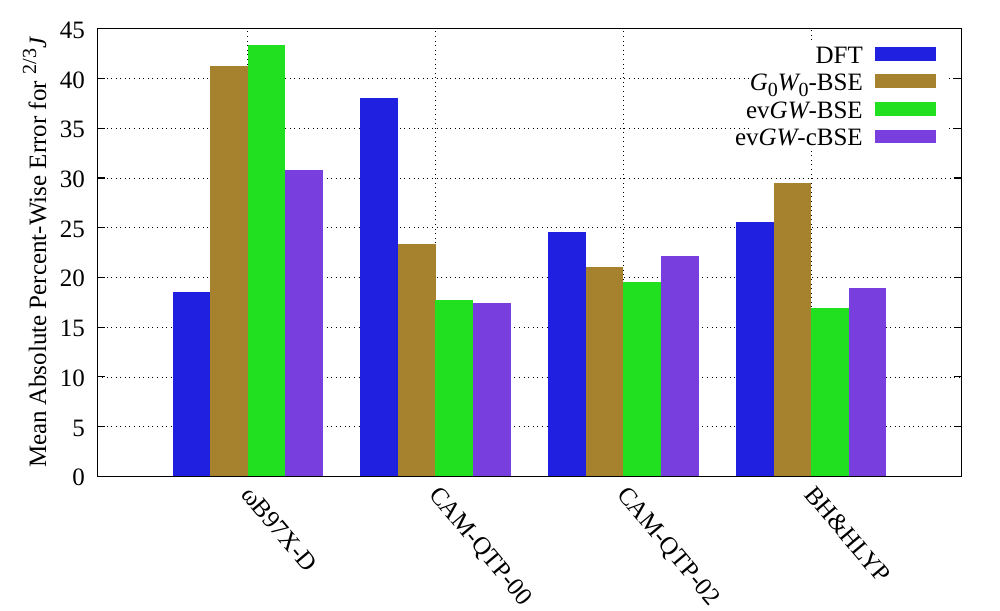}
    \caption{Mean absolute percent-wise error for NMR $^1J$ coupling constants
    (left panel) and NMR $^{2/3}J$ coupling constants (right panel) with respect
    to the CC3/aug-ccJ-pVTZ reference values of Ref.~\citenum{Faber.Sauer.ea:Importance.2017}. 
    {\textomega}B97X-D is one of the top performers for Kohn--Sham DFT.
    Results are taken from Refs.~\citenum{Holzer.Franzke.ea:Assessing.2021}
    (Kohn--Sham DFT) and \citenum{Franzke.Holzer.ea:NMR.2022} ($GW$-BSE).
    The test set consists of 13 molecules and 45 NMR chemically inequivalent
    coupling constants.}
    \label{fig:sscc-benchmark}
\end{figure*}

A first benchmark study was performed in Ref.~\citenum{Franzke.Holzer.ea:NMR.2022}
for a set of organic molecules with CC3 serving as reference and results were
compared to Kohn--Sham DFT.
Often, the FC interaction is the dominant contribution and the $GW$-BSE does not
improve upon DFT. Due to the relation of the FC term to triplet excitations, this
property is a serious challenge for $GW$-BSE as illustrated in Figure~\ref{fig:sscc-benchmark}.
Generally, benchmark studies have shown that $GW$-BSE performs much better
for singlet excitations than triplet excitations.
\cite{Jacquemin.Duchemin.ea:Benchmark.2017, Rangel.Hamed.ea:assessment.2017,
Gui.Holzer.ea:Accuracy.2018}
Therefore, the correlation-kernel augmented Bethe--Salpeter equation (cBSE)
was introduced in Ref.~\citenum{Holzer.Klopper:Communication.2018}, which
computes the screened exchange with the Kohn--Sham orbitals and includes
the correlation part of the XC kernel in the electronic Hessian of the BSE.
This improves the description of triplet excitations, while retaining the
correct description of charge-transfer processes.
According to Figure~\ref{fig:sscc-benchmark}, cBSE also improves the accuracy
of NMR coupling constants. As no qs$GW$ is used, the underlying functional
has to incorporate a large amount of exact exchange such as in Becke's
half and half functional (BH\&HLYP) or the CAM-QTP family which was specifically
designed to yields good ionization energies.

\begin{figure}[htbp]
  \includegraphics[width=\columnwidth]{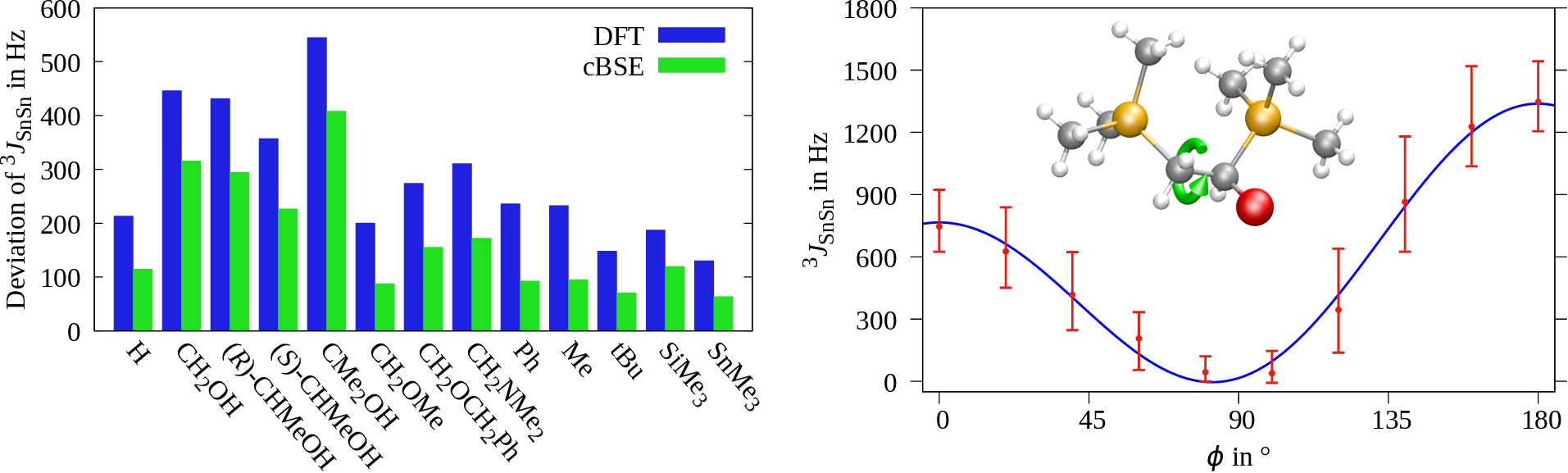}
  \caption{Left panel: Deviations of the Boltzmann-averaged coupling constants of the 13 tin
  compounds. 
  Deviations with the BH{\&}HLYP functional are shown in blue and those with ev$GW$-cBSE{\makeatletter @}BH{\&}HLYP in green.
  Abbreviations: Me = methyl, Ph = phenyl, tBu = tert-butyl, \textit{S} = sinister,
  \textit{R} = rectus.
  Right panel: For each tin--tin torsion angle $\phi$, the average of the
  $^3J_{\text{SnSn}}$ coupling constant over the 13 compounds is given (red dot)
  as well as the total range spanned (vertical bars) at the ev$GW$-cBSE{\makeatletter @}BH{\&}HLYP/x2cTZVPall-2c level.
  A fitted Karplus equation is given in blue. Fit parameters for Eq.~\ref{eq:karplus}
  are $A = 518 $\,Hz, $B = -286 $\,Hz, $ C= 534$\,Hz. The torsion angle and the molecular
  structure are illustrated above the fit. Color code: white hydrogen, gray carbon, golden tin,
  red subsituent R. Reproduced (adapted) from Y.\ J.\ Franzke, C.\ Holzer,
  F.\ Mack, \textit{J.\ Chem.\ Theory Comput.}\ \textbf{2022}, \textit{18}, 1030--1045.
  Copyright the Authors. Published by American Chemical Society.}
  \label{fig:karplus}
\end{figure}

The excellent performance of the ev$GW$-cBSE based on a BH\&HLYP starting point
is not restricted to small systems. Notably, it can be readily applied to
study the Karplus curve of tin compounds at the relativistic two-component level,
which treats scalar-relativistic and spin--orbit effects on an equal footing.
According to the Karplus equation, the NMR $^3J$ coupling constant of
the molecules (CH$_3$)$_3$Sn--CH$_2$--CHR--SnMe$_3$, with R being different
substituents, follows the relation
\begin{equation}
\label{eq:karplus}
^3J (\phi) = A \cos(2\phi) + B \cos(\phi) + C
\end{equation}
where $\phi$ denotes the torsion angle and $A, B, C$ are constants or
fit parameters. To study this relation and the accuracy of the BSE
framework for larger systems, 13 different substituents ranging from
a simple hydrogen atom to the trimethylstannyl group
are used and the torsion angles are varied from 0 to 180$^{\circ}$.
First, a Boltzmann-average then allows to compare the calculated coupling
constants to the experimental findings \cite{Mitchell.Kowall:Karplus-type.1995}
for each substituent as shown in the left panel of Figure~\ref{fig:karplus}.
Here, the cBSE substantially reduces the deviation from the experiment compared
to the conventional DFT approach.
Secondly, an average over the 13 compounds at each angle allows to plot
the coupling constants vs.\ the torsion angle and fit the results to
Eq.~\ref{eq:karplus}.
As indicated by a coefficient of determination of $R^2 = 0.99$, this
fit to the Karplus relation works excellently. That is, cBSE successfully accounts
for the triplet inaccuracies of the original BSE approach.

Taking together, the BSE can be successfully applied beyond
light-matter interactions. The computational protocol for NMR
coupling constants relies on ev$GW$ and a suitable functional
approximation. This dependence on the Kohn--Sham starting point
could be mitigated by the qs$GW$ approach. Unfortunately, qs$GW$
is often plagued by convergence issues and comes with increased
computational costs.

\section{Ground-State Correlation Energies and General Properties}
We finally note that another important step in the direction of ground-state
molecular properties was taken in Refs.~\citenum{Holzer.Gui.ea:Bethe-Salpeter.2018}
and \citenum{Loos.Scemama.ea:Pros.2020}, where the authors calculated
the BSE ground-state correlation energy of small molecules with the
adiabatic-connection fluctuation-dissipation theorem.
\cite{Maggio.Kresse:Correlation.2016}
Bond lengths, potential energy curves, and vibrational frequencies
were subsequently obtained numerically. Here, the BSE correlation
energy is given by an integration over the coupling strength parameter $\lambda$
and reads \cite{Maggio.Kresse:Correlation.2016, Holzer.Gui.ea:Bethe-Salpeter.2018}
\begin{equation}
\label{eq:bse-corr}
E_{\text{c}} = \frac{1}{2} \int_0^1 \text{d} \lambda ~ \Tr \left(\textbf{C} \textbf{D}_{\lambda}\right)
\end{equation}
with the interaction kernel \textbf{C} at full coupling strength ($\lambda = 1$)
defined as
\begin{equation}
    \textbf{C} = \textbf{C} _{\lambda = 1} =
    \begin{pmatrix}
        \Tilde{\textbf{A}} & \Tilde{\textbf{B}} \\
        \Tilde{\textbf{B}}^* & \Tilde{\textbf{A}}^*
    \end{pmatrix}
\end{equation}
Here, the matrices $\Tilde{\textbf{A}}$ and $\Tilde{\textbf{B}}$
are a simplified form of the BSE quantities,
\begin{align}
\label{eq:Atilde}
    \tilde{A}_{ai,bj} = & (ai|jb)  \\
\label{eq:Btilde}
    \tilde{B}_{ai,bj} = & (ai|bj)
\end{align}
Eqs.~\ref{eq:Atilde} and \ref{eq:Btilde} resemble the interaction kernel 
of the direct random phase approximation (RPA), which illustrates the 
relationship of the RPA and $GW$-BSE methods.
\textbf{D}$_{\lambda}$ is the correlation part of the two-electron density
matrix at a given coupling strength
\begin{equation}
\textbf{D}_{\lambda} =
\begin{pmatrix}
\textbf{Y} _{\lambda} \textbf{Y} _{\lambda}^{\dag}  & \textbf{Y} _{\lambda} \textbf{X} _{\lambda}^{\dag} \\
\textbf{X} _{\lambda} \textbf{Y} _{\lambda}^{\dag}  & \textbf{X} _{\lambda} \textbf{X} _{\lambda}^{\dag}
\end{pmatrix}^*
-
\begin{pmatrix}
    \mathbf{0} & \mathbf{0} \\
    \mathbf{0} & \mathbf{1} 
\end{pmatrix}
\end{equation}
where the vectors \textbf{X} and \textbf{Y} are obtained by solving the general 
problem described by Eq.~\ref{eq:fullBSE} with the accordingly modified matrix
elements of \textbf{A}$_{\lambda}$ and \textbf{B}$_{\lambda}$.
The subscript $\lambda$ indicates that the two-electron integrals in Eqs.~\ref{eq:Afull}
and \ref{eq:Bfull} are scaled by the coupling strength parameter $\lambda$.
We stress that both \textbf{C} and \textbf{D}$_{\lambda}$ are Hermitian.
As usual, the equations can be simplified for closed-shell systems and real-valued
orbitals as shown in Ref.~\citenum{Loos.Scemama.ea:Pros.2020}.
The total energy is calculated by adding the nuclear repulsion energy
and the electronic Hartree--Fock energy to the obtained correlation energy.
Formally, the correlation energy in Eq.~\ref{eq:bse-corr} may be used to
construct a Lagrangian similar to the application of the random-phase
approximation to molecular properties in analytical derivative theory.
\cite{Rekkedal.Coriani.ea:Communication.2013}
Unfortunately, the calculation of the BSE correlation energy is still
plagued by stability issues when he regular BSE expressions for the
matrices \textbf{A} and \textbf{B} are applied and simply scaled with
$\lambda$. This can be resolved by completely deriving the screened
exchange at a given coupling strength. \cite{Holzer.Gui.ea:Bethe-Salpeter.2018}
Still, unphysical irregularities are observed for the ground-state
potential energy surfaces which are due to the quasiparticle energies.
In detail, these issues arise from discontinuities of the $GW$ quasiparticle
energies as a function of the interatomic distance or bond length. 
\cite{Loos.Scemama.ea:Pros.2020}
This issue limits the applicability of Eq.~\ref{eq:bse-corr} in practice
and may prevent structure optimizations for both ground states and
excited states.

A special case of a correlation energy obtained from the BSE is the 
intermolecular dispersion energy, that can be obtained from symmetry
adapted perturbation theory (SAPT). It can be evaluated from the
an integration over the imaginary axis of the dipole polarizability
of the two molecular systems $\mathbf{1}$ and $\mathbf{2}$, 
\cite{Holzer.Klopper:Communication.2017}
\begin{equation}
    E_{\text{c}}^{\text{disp.}} = \frac{1}{2\pi} \int_0^{\infty} \text{d}\omega \sum_{ai,bj}
    \gamma_{ai}^{\alpha,\mathbf{1}}(\text{i}\omega) \gamma_{bj}^{\alpha,\mathbf{2}}(\text{i}\omega) (ai|bj) 
\end{equation}
The quality of dispersion energies calculated from the BSE is therefore
tightly linked to the quality of polarizabilities.
As outlined earlier in Table~\ref{tab:polarizabilities}, 
these can assumed to be very high. Accordingly, 
Ref.~\citenum{Holzer.Klopper:Communication.2017} found
high quality molecular interaction energies for van der Waals-bonded systems, 
clearly exceeding the quality of state-of-the-art DFT based prediction
as outlined in Table~\ref{tab:bzimpy}. For the weakly bonded benzene-imidazole
and benzene-pyrrole systems, both $G_0W_0$-BSE and  ev$GW$-BSE based SAPT 
only deviate by a fraction of a kJ/mol from reference coupled cluster
singles, doubles, and perturbative triples (CCSD(T)) values.
The basis set dependence for these calculations is more pronounced than
the dependence on the quasiparticle energies, as the deviation between
results from $G_0W_0$ and ev$GW$ is smaller than the deviation between results
from different basis sets or the correction from the explicitly correlated F12
ansatz. As CCSD(T) with large basis sets is considered the gold standard for
van der Waals-bonded systems, this hints at correlation energies from the BSE
being very accurate in certain cases.

\begin{table}[ht!]
 \caption{Electronic binding energy $D_e$ (in kJ/mol) of benzene-imidazole (Bz$\cdot$Im)
 and benzene-pyrrole (Bz$\cdot$Py) calculated at the SAPT(Method)/aug-cc-pV(D/T/Q)Z 
 level of theory. $GW$-BSE results calculated at the PBE0 Kohn--Sham reference. 
 Results are taken from Ref.~\citenum{Holzer.Klopper:Communication.2017}.}
 \label{tab:bzimpy}
\begin{tabular}{@{\extracolsep{6pt}}
l
S[table-format = 2.1]
S[table-format = 2.1]
S[table-format = 2.1]
S[table-format = 2.1]
S[table-format = 2.1]
S[table-format = 2.1]
@{}
}
\toprule
 & \multicolumn{2}{c}{aug-cc-pVDZ} & \multicolumn{2}{c}{aug-cc-pVTZ} & \multicolumn{2}{c}{aug-cc-pVQZ} \\
 \cmidrule{2-3} \cmidrule{4-5} \cmidrule{5-7}
{\small Method} & Bz$\cdot$Im & Bz$\cdot$Py & Bz$\cdot$Im & Bz$\cdot$Py & Bz$\cdot$Im & Bz$\cdot$Py \\ 
\midrule
\small PBE0AC        & 24.2 & 21.5 & 23.1 & 20.8 & \text{--} & \text{--}  \\
\small +F12          & 26.1 & 23.4 & 23.8 & 21.4 & \text{--} & \text{--}  \\ \midrule
\small $G_0W_0$-BSE  & 20.5 & 18.3 & 22.1 & 19.9 & 22.4 & 20.2 \\
\small +F12          & 22.5 & 20.2 & 22.8 & 20.5 & 22.7 & 20.5 \\ \midrule
\small ev$GW$-BSE    & 20.1 & 17.9 & 21.5 & 19.4 & \text{--} & \text{--} \\ 
\small +F12          & 22.0 & 19.8 & 22.2 & 20.0 & \text{--} & \text{--} \\ \midrule
\small CCSD(T)       & 19.6 & 17.5 & 21.9 & 19.7 & \text{--} & \text{--} \\
\small +(F12)(T*)    & 22.3 & 20.0 & 22.8 & 20.4 & 22.8 & 20.5 \\
\bottomrule
\end{tabular}
\end{table}

\section{Summary and Possible Future Directions}
\label{sec:summary}

This review outlines that the Bethe--Salpeter equation method has well advanced 
past the point of mainly being useful for simulating linear optical spectra.
During the last few years, methods to also simulate various properties, 
including polarizabilities, hyperpolarizabilities, two-photon absorption, 
transient absorption, optical rotation, NMR properties, and properties linked with 
these have emerged. Properties evaluated from the BSE have been shown to be very reliable, 
often outperforming TD-DFT at the same computational scaling.
For example, the general weakness of TD-DFT, frequently being unable to
properly describe charge-transfer or Rydberg excitations is cured by the BSE, 
in turn also leading to an improved description of polarizabilities
and other properties. And while TD-DFT certainly remains competitive, 
the ever-growing number of functionals are both a blessing and a
curse---a blessing because there will likely be at least one density functional 
that works, and a curse because it gets evermore challenging to identify
the correct one. The BSE method outlined in this review removes the 
challenge of finding the correct functional, replacing it with a more 
rigorous approach from first principles. Most of the dependence on the
underlying functional is removed by the preceding $GW$ step, and results
from the  combined $GW$-BSE approach are therefore robust. This is an
incredibly  valuable feature, and we therefore expect the BSE to be a growing
competition to TD-DFT on the prediction of properties over the next decade.

Finally, a current trend in calculating correlation energies from the BSE
has emerged and proven useful to study the underlying physics
of many-electron systems. The BSE is there used to restore the 
electron-electron interaction, leading to an improved description of
the fluctations taking place in this delicate process.
Further, the $GW$-BSE can even be applied to a many-Fermions framework going
beyond electrons within a multicomponent ansatz to compute excitation
energies. When using a common Hilbert space for the RI auxiliary basis sets,
the working equations for the $GW$ quasiparticles can be derived in a
straightforward way and used to set up the BSE Hessian matrices.
\cite{Holzer.Franzke:Beyond.2024} This suggests that also response
properties could become available for this framework in the near
future.

\section*{Data Availability Statement}
No new data were generated or analyzed in support of this study.

\section*{Author Contributions and Declarations}
\noindent \textbf{Christof Holzer}: Conceptualization (equal);
Investigation (equal); Methodology (equal);
Validation (equal); Visualization (equal);
Writing – original draft (equal); Writing – review \& editing (equal).
 \\
\noindent \textbf{Yannick J. Franzke}: Conceptualization (equal);
Investigation (equal); Methodology (equal);
Validation (equal); Visualization (equal);
Writing – original draft (equal); Writing – review \& editing (equal).
\medskip \newline
\noindent \textbf{Notes} \\
\noindent The authors declare no competing financial interest.

\begin{acknowledgement}
C.H. gratefully acknowledges the Volkswagen Stiftung for financial
support.
Y.J.F.\ gratefully acknowledges support via the Walter--Benjamin
programme funded by the Deutsche Forschungsgemeinschaft (DFG, German
Research Foundation) --- 518707327.
\end{acknowledgement}

\bibliography{literature}

\end{document}